\documentstyle[preprint,tighten,eqsecnum,floats,aps,epsfig]{revtex}

\begin{document}

\preprint{nucl-th/9610005}
\title{On the Coulomb and higher-order sum rules in the relativistic Fermi gas
}
\author{
 {P. Amore}$^a$, {R. Cenni}$^b$, {T. W. Donnelly}$^c$ and {A. Molinari}$^a$
\footnote[1]{This work is supported in part by funds provided by the U.S.
Department of Energy (D.O.E.) under cooperative agreement
\#DE-FC01-94ER40818.
\vskip0.5truecm
MIT/CTP \#2576}}
\address{
 $^a$ Dipartimento di Fisica Teorica dell'Universit\`a
 di Torino \\
 Istituto Nazionale di Fisica Nucleare, Sezione di Torino, \\
 via P.Giuria 1, I-10125 Torino (Italy) \\
 $^b$ Dipartimento di Fisica dell'Universit\`a di Genova \\
 Istituto Nazionale di Fisica Nucleare, Sezione di Genova, \\
 via Dodecaneso, 33 - 16146 Genova (Italy)\\
 Center for Theoretical Physics\\
 $^c$ Laboratory for Nuclear Science and Department of Physics\\
 Massachusetts Institute of Technology\\
 Cambridge, Massachusetts 02139, U. S. A.\\~
}
\date{\today}

\maketitle

\begin{abstract}
Two different methods for establishing a space-like
Coulomb sum rule for the relativistic Fermi gas are 
compared. Both of them divide the charge response by a normalizing
factor such that the reduced response thus obtained fulfills the sum rule
at large momentum transfer. To determine the factor, in the first approach 
one exploits
the scaling property of the longitudinal response function, while
in the second one
enforces the completeness of the states in the space-like domain via the 
Foldy-Wouthuysen transformation.
The energy-weighted and the squared-energy-weighted  sum rules for the reduced
responses are explored as well and the extension to momentum distributions 
that are more
general than a step-function is also considered.
The two methods yield reduced responses and Coulomb sum rules that 
saturate in the
non-Pauli-blocked region, which can hardly be distinguished for Fermi momenta
appropriate to atomic nuclei. Notably the sum rule obtained in the
Foldy-Wouthuysen approach coincides with the well known non-relativistic one.
Only at quite large momentum transfers (say 1 GeV/c) does 
a modest softening of the
Foldy-Wouthuysen reduced response with respect to that obtained in
the scaling framework show up. The two responses have the same
half-width to second order in the Fermi momentum expansion.
However, when distributions extending to momenta larger
than that at the Fermi surface are employed, then in both methods the Coulomb sum rule
saturates only
if the normalizing factors are appropriately modified to account for the high
momentum components of the nucleons. 
\end{abstract}

\vfill
\begin{center}
Submitted to: {\em Nuclear Physics A}
\end{center}

\vfill
\hfill mmmmm 1996

\eject

\section{Introduction}

In this paper we compare two different approaches to the relativistic Coulomb 
sum rule  confining ourselves to dealing with a homogeneous, translationally 
invariant system of non-interacting nucleons, namely the Fermi gas (FG).
We consider not only the customary step-function, but more 
general momentum distributions as well in order to reflect some of 
the correlations among the nucleons. Our chief aim is to explore whether the
relativistic Coulomb sum rule displays the same basic feature 
as the non-relativistic one, i.e. saturation at large transferred momentum.
Indeed, as is well-known, for a FG with $Z$ non-interacting protons the 
non-relativistic Coulomb sum rule (NRCSR) reads ($k_F$ is the Fermi momentum)
\begin{eqnarray}
\int_0^{\infty} \frac{R_L (\bbox{q},\omega)}{Z G_E^2 (Q^2)} d\omega &=&
\int_0^{\infty} r_L(\bbox{q},\omega) d\omega = 
\Sigma_C^{nr} (q)  \nonumber \\
&=&  \left[ \vartheta(q-2 k_F) + \frac{3 q}{4 k_F}
\left(1-\frac{q^2}{12 k_F^2}\right) \vartheta(2 k_F-q) \right]
\label{eq:intro1}
\end{eqnarray}
and exhibits saturation as a consequence of unitarity (summation over a
complete set of states) in the non-Pauli-blocked regime. 
Here the NRCSR just counts the charged particles inside the system.
In (\ref{eq:intro1}) $R_L$ is the usual longitudinal response function whereas
$r_L$ is commonly referred to as the reduced longitudinal 
response function, since its dependence upon the physics of the nucleon
has been divided out ($G_E(Q^2)$ is the electric form factor, here
 of the proton).
Note that the implementation of unitarity in the non-relativistic regime 
requires extending the range of integration in (\ref{eq:intro1}) up to
infinity.

Now from the well-known symmetry property of $R_L$ in the Pauli-blocked 
region\cite{Ros80,Cen88}, namely
\begin{eqnarray}
R_L^{b} (q,\omega) = R_L^{nb} (q,\omega) - R_L^{nb} (q,-\omega) \ \ \ ,
\label{eq:simmetria}
\end{eqnarray}
the relationship
\begin{eqnarray}
\int_0^{\infty} \frac{R_L^{nb} (\bbox{q},\omega)}{Z G_E^2 (Q^2)} d\omega  = 
\int_0^{\infty} r_L^{nb}(q,\omega) d\omega =
\frac{1}{2} \left[ 1 + \Sigma_C^{nr} (q) \right] 
\label{eq:intro2}
\end{eqnarray}
may be deduced (the superscript b(nb) 
stands for Pauli(non-Pauli)-blocked), which
allows one to express the NRCSR {\it solely} in terms of the non-Pauli-blocked 
reduced response over the whole range of $q$. 

When attempting to generalize the NRCSR to a relativistic homogeneous system 
with an equal number of non-interacting protons and neutrons (the symmetric
relativistic Fermi gas (RFG)) one faces at least two issues in 
order to achieve saturation: first the neutrons and protons composite nature, 
which in the relativistic domain is not so straightforward to factor out, 
should be accounted for; moreover the closure relation is no longer
restricted to particle-hole ($ph$) excitations, 
but includes the time-like region (particle-antiparticle 
($p\overline{p}$) excitations) as well.

For an ideal system of pointlike non-interacting nucleons without 
anomalous magnetic moments the first of the above items is of course avoided,
whereas the second, as shown by Walecka\cite{Wal83} and Matsui\cite{Mat83},
can be dealt with without difficulty: a relativistic 
Coulomb sum rule (RCSR) that is
quite close to the NRCSR is thus obtained for densities of the RFG roughly 
corresponding to those in nuclei by exploiting closure in 
both space-like and time-like regions, i.e. by integrating over the whole range
of positive energies (of course for a given $q$ the actual range of 
integration is cut by the sharp boundaries of the FG response function).

However experimentally the structure of the nucleon should be reckoned with and
the time-like energy domain cannot be reached with the presently available
experimental facilities.
Therefore one would like to define a sum rule where the physics of the nucleon
has been disentangled (at least to a large extent) and saturation obtains in 
the space-like energy regime alone.
For this purpose a method referred to as the {\it scaling approach} has
been developed by Alberico et al.\cite{Alb93}. In the present paper
another one, based on the
Foldy-Wouthuysen (FW) transformation, is suggested and compared with the 
scaling method not only in
connection with the Coulomb sum rule, but also with those having
 energy and  energy squared weightings.

\section{}
\subsection{The scaling approach}
It is well-known that in the non-relativistic FG for large enough
transferred momenta, namely for $q>2 k_F$, the reduced response $r_L$ scales, 
i.e. it becomes function of only one variable, the so-called scaling variable, 
which corresponds to the minimum momentum  parallel or antiparallel
to $\bbox{q}$ that a nucleon inside the system can have in order to 
contribute to the response.
A dimensionless non-relativistic scaling variable can indeed be defined 
as\cite{Wes75}\footnote{Following ref. \cite{Alb88} we introduce here
the dimensionless 
momentum and energy transfer to the nucleus, the
momentum and energy of a nucleon 
inside the RFG and the momentum and Fermi energy according to: \\ 
$\kappa \equiv q/{2 m_N}$, $\lambda \equiv \omega/{2 m_N}$ 
$\eta \equiv k/m_N$, $\epsilon \equiv \sqrt{1+\eta^2}$,
$\eta_F \equiv k_F/m_N$, $\epsilon_F \equiv \sqrt{1+\eta_F^2}$. \\
Moreover  $\tau = \kappa^2-\lambda^2$ and $\xi_F \equiv \epsilon_F-1$.}
\begin{equation}
\psi_{nr} \equiv \frac{k_{\parallel}}{m_N} = 
\frac{m_N}{k_F} \left(\frac{\omega}{q}-\frac{q}{2 m_N}\right) =
\frac{1}{\eta_F} \left(\frac{\lambda}{\kappa}-\kappa\right) \ \ \ ,
\end{equation}
where $m_N$ is the nucleon mass,
in terms of which the non-relativistic reduced longitudinal response reads
\begin{eqnarray}
r_L^{nrFG}(\kappa,\lambda) &=& \frac{3 }{4 m_N \kappa \eta_F^3} 
\left\{\vartheta\left(\eta_F-\kappa\right) 
\left[ \frac{\eta_F^2}{2} (1-\psi_{nr}^2) \vartheta(1-\psi_{nr}) 
\vartheta\left(\frac{\kappa}{\eta_F}-\frac{1-\psi_{nr}}{2}\right) + \right. 
\right. \nonumber \\
&& \left. \left. + 2 \lambda
\vartheta\left(\frac{1-\psi_{nr}}{2}-\frac{\kappa}{\eta_F}
\right)
\right] + \frac{\eta_F^2}{2} (1-\psi_{nr}^2) 
\vartheta(1-\psi_{nr}^2) \vartheta\left(\kappa-\eta_F\right) \right\} \ \ \ .
\end{eqnarray}
Thus, as an alternative to (\ref{eq:intro1}),
one can obtain the NRCSR as an integral over the scaling variable.
Indeed it is easily verified that
\begin{eqnarray}
\int_{-1}^{+1} 2 m_N r_L^{nrFG}(\kappa,\lambda)
\frac{\partial\lambda}{\partial \psi_{nr}} d\psi_{nr} = \Sigma_C^{nr} (\kappa) 
\ \ \ .
\end{eqnarray}

Also for the RFG, a scaling variable can be defined, again representing 
the minimum longitudinal momentum of the nucleon absorbing
the virtual photon inside nuclear matter. It reads
\cite{Alb88}
\begin{eqnarray}
\psi \equiv \sqrt{\frac{\gamma_{-} -1}{\xi_F}}
\left[ \vartheta(\lambda-\lambda_0) - \vartheta(\lambda_0-\lambda) \right]
\label{eq:scal0}\;,
\end{eqnarray}
where $\lambda_0 = \frac{1}{2} \left[ \sqrt{1+4 \kappa^2} - 1 \right]$ is
the dimensionless quasielastic peak energy and $\gamma_{-} \equiv \kappa 
\sqrt{1+\frac{1}{\tau}} - \lambda$. 
In terms of the above the space-like longitudinal response function is then
expressed as follows \cite{Alb88}
\begin{eqnarray}
R_L^{RFG} (\kappa,\lambda) &=& \frac{3 {\cal N}}{4 m_N \kappa \eta_F^3} 
\left(\epsilon_F - \Gamma\right) \vartheta \left(\epsilon_F-\Gamma\right) 
U_L(\kappa,\lambda) \nonumber \\
&=& \frac{H_L(\kappa,\lambda)}{2 m_N} \frac{\partial \psi}{\partial \lambda}
\frac{3}{4} \vartheta\left(\epsilon_F - \Gamma \right)
\left\{ \vartheta\left(\eta_F-\kappa\right) 
\left[ (1-\psi^2) \vartheta\left(\kappa \sqrt{1+\frac{1}{\tau}} + 
\lambda - \epsilon_F \right)  + \right. \right. \nonumber \\
&& \left. \left. + \frac{2 \lambda}{\xi_F}
\vartheta\left(\epsilon_F - \kappa \sqrt{1+\frac{1}{\tau}} - \lambda \right)
\right] +\vartheta\left(\kappa-\eta_F\right) (1-\psi^2) 
  \right\} \ \ \ ,
\label{eq:scal1}
\end{eqnarray}
and is non-vanishing in the range $\lambda_{min} < \lambda < 
\lambda_{max}$, where 
\begin{eqnarray}
\lambda_{max,min} \equiv \frac{1}{2} \left[ \sqrt{1+ (2 \kappa \pm \eta_F)^2} -
\epsilon_F \right] \ \ \ .
\end{eqnarray}
In  (\ref{eq:scal1}) 
\begin{eqnarray}
U_L(\kappa,\lambda) = \frac{\kappa^2}{\tau} \left[G_E^2(\tau) + W_2(\tau)\Delta
\right]
\end{eqnarray}
and
\begin{eqnarray}
H_L^{RFG}(\kappa,\lambda) = \frac{2 \xi_F}{\kappa \eta_F^3} \frac{{\cal N}}
{\frac{\partial \psi}
{\partial \lambda}} U_L(\kappa,\lambda) \equiv \frac{2 \xi_F}{\kappa \eta_F^3} 
\frac{{\cal N}}{\frac{\partial \psi}{\partial \lambda}} \frac{\kappa^2}{\tau} 
\left[G_E^2 (\tau) + W_2(\tau) \Delta \right] \ \ \  
\label{eq:cenni}
\end{eqnarray}
with $W_2(\tau) = \frac{1}{1+\tau}\left(G_E^2(\tau)+\tau G_M^2(\tau)
\right)$, $G_M (\tau)$ being the magnetic form factor of the nucleon.
Furthermore
\begin{eqnarray}
\Delta &\equiv& \frac{\langle k_{\perp}^2 \rangle}{m_N^2} = 
\frac{1}{\epsilon_F-\Gamma} \int_{\Gamma}^{\epsilon_F} d\epsilon 
\eta_{\perp}^2 (\epsilon,\kappa,\lambda) = \nonumber \\
&=& \frac{\tau}{\kappa^2}
\left[ \frac{\left(\epsilon_F^2+\Gamma \epsilon_F +\Gamma^2\right)}{3} + \lambda
(\epsilon_F+\Gamma) + \lambda^2\right] - \left(1+\tau\right)  ,
\label{eq:scal2}
\end{eqnarray}
where $\Gamma \equiv {\rm max}\left[ \epsilon_F-2 \lambda, \gamma_{-}\right]$ 
and $\eta_{\perp}^2 (\epsilon,\kappa,\lambda) = \frac{\tau}{\kappa^2}
(\epsilon+\lambda)^2-(1+\tau)$  corresponds to the average quadratic
transverse momentum in units of $m_N^2$. Indeed the motion of the nucleons
transverse to $\bbox{q}$ introduces a {\it magnetic} contribution into the
{\it charge} response\cite{Cha93}.

It should be understood that the charge response is calculated by adding 
the contribution with neutrons where ${\cal N} = N$ and 
with protons where ${\cal N} = Z$.

The structure of (\ref{eq:scal1}) naturally suggests introducing a reduced
response according to 
\begin{eqnarray}
r_L^{RFG} (q,\omega) = \frac{R_L^{RFG}(q,\omega)}{H_L^{RFG}(q,\omega)} \ \ \ .
\end{eqnarray}
For $\kappa>\eta_F$, where Pauli correlations are no longer
effective, this reads
\begin{eqnarray}
r_L (q,\omega) = \frac{3}{4} \frac{1}{2 m_N} \left(1-\psi^2\right) 
\vartheta\left(1-\psi^2\right) \frac{\partial \psi}{\partial \lambda} = S(\psi) 
\frac{\partial \psi}{\partial \lambda} 
\end{eqnarray}
and is indeed well suited to be integrated over $\psi$, since it
scales\footnote{In fact 
$S(\psi)$ is proportional to the scaling function defined in \cite{Alb88}}
and has the Jacobian incorporated.

The RCSR for the reduced longitudinal response in the non-Pauli-blocked 
region ($\kappa > \eta_F$) is thus easily obtained\cite{Cen96} according to
\begin{eqnarray}
\Sigma_C(q) =
\int_0^{q} d\omega\frac{R_L^{RFG} (q,\omega)}{H_L^{RFG}(\kappa,\lambda)} = 
\int_0^{\kappa} d\lambda \frac{d\psi}{d\lambda} S(\psi)=
\int_{-1}^{+1} d\psi S(\psi) = 1 \ \ \ ,
\end{eqnarray}
where the range of integration is cut on the light front, since
no anti-nucleon physics is contained in our $R_L$: 
therefore $\Sigma_C(q)$ is directly accessible to the experiment.
Because of the scaling it thus appears that it is possible to define for the
RFG  a {\it space-like} reduced response fulfilling a {\it space-like} Coulomb
sum rule which saturates when the Pauli correlations vanish. 

Although the RFG is not trustworthy as a model for nuclei at small
$q$ (here the surface matters), one would still like to
get a complete analytical expression for $\Sigma_C(q)$ as obtained via the
scaling method in the RFG. This turns out to be extremely cumbersome for
$q\leq 2k_F$.
However (\ref{eq:intro2}), although no longer ``exactly'' valid, 
still provides an approximate expression  for the RFG Coulomb sum rule in the
Pauli-blocked domain, which is quite accurate in the range of densities
appropriate for nuclei (see Fig.~\ref{fig:scal}). It reads
\begin{eqnarray}
\Sigma_C (q) \approx \left[2 \Sigma_C (q)^{nb} -1 \right] = \frac{3}{2}
\sqrt{\frac{\sqrt{1+\kappa^2}-1}{\xi_F}} \left[ 1 - \frac{1}{3 \xi_F} \left(
\sqrt{1+\kappa^2}-1\right)\right]
\label{eq:scal3}
\end{eqnarray}
and reduces to the non-relativistic Pauli-blocked Coulomb sum rule
$\Sigma_C^{nr}$ in the small momentum, small density limit.

\subsection{The Foldy-Wouthuysen approach}

We shall now attempt to recover the saturation value for the Coulomb sum
rule by exploiting the FW formalism, which has been 
previously successfully used in a potential model description of nuclear 
matter \cite{ABD95}.
The FW framework yields an expression for the response function 
which is convenient for the non-relativistic reduction, although it does not
transparently display covariance.

Through the unitary transformation
\begin{equation}
    {\cal T} (\bbox{k}) = \sqrt{\frac{E_{k} + m_N}{2 E_{k}}} 
     \left( \openone + \frac{\bbox{\gamma} \cdot \bbox{k}}{E_{k} 
     + m_N} \right)  \ 
\label{eq:fol1}
\end{equation}
the FW reduced Green's function is defined as follows
\begin{eqnarray}
  G^{\text{FW}} (k) = {\cal T}(\bbox{k}) G (k) {\cal T}(\bbox{k}) \ \ \ 
,
\label{fol2}
\end{eqnarray}
where, in the non-interacting case\cite{Ser86,Cen94},
\begin{eqnarray}
  G(k) = G_0 (k) = \frac{\left(\rlap/k + m\right)}{2 E_{k}} \left[ 
\frac{\vartheta(k-k_F)}{k_0-E_{k}+i\varepsilon} + 
\frac{\vartheta (k_F - k)}{k_0-E_{k}-i \varepsilon} -
\frac{1}{k_0 + E_{k} - i \varepsilon} \right] 
\label{fol2bis}
\end{eqnarray}
with $E_{k}= \sqrt{k^2 + m_N^2}$, i.e. the free relativistic energy.
Owing to the identity
\begin{eqnarray}
{\cal T}(\bbox{k}) \left(\gamma^{\mu} k_{\mu} + m\right) {\cal T}(\bbox{k}) = 
P_{+} (k_0+E_k) - P_{-} (k_0-E_k)  \ 
\end{eqnarray}
one then gets
\begin{eqnarray}
   G_0^{\text{FW}} (k) &=& P_{+} \left[ {\frac{\vartheta (k-k_F)}{k_0 -
 E_{k} + i \varepsilon}} + {\frac{\vartheta (k_F - k)}{k_0 -
 E_{k} - i \varepsilon}} \right] - {P_{-}} {\frac{1}{k_0 +
 E_{k} - i \varepsilon}} \ ,
\label{eq:fol3}
\end{eqnarray}
where the operators $P_{\pm} =  {\frac{\openone \pm \gamma^0}{2}}$ 
project  on the large/small components of the wave function.

\begin{figure}
\begin{center}
\mbox{\epsfig{file=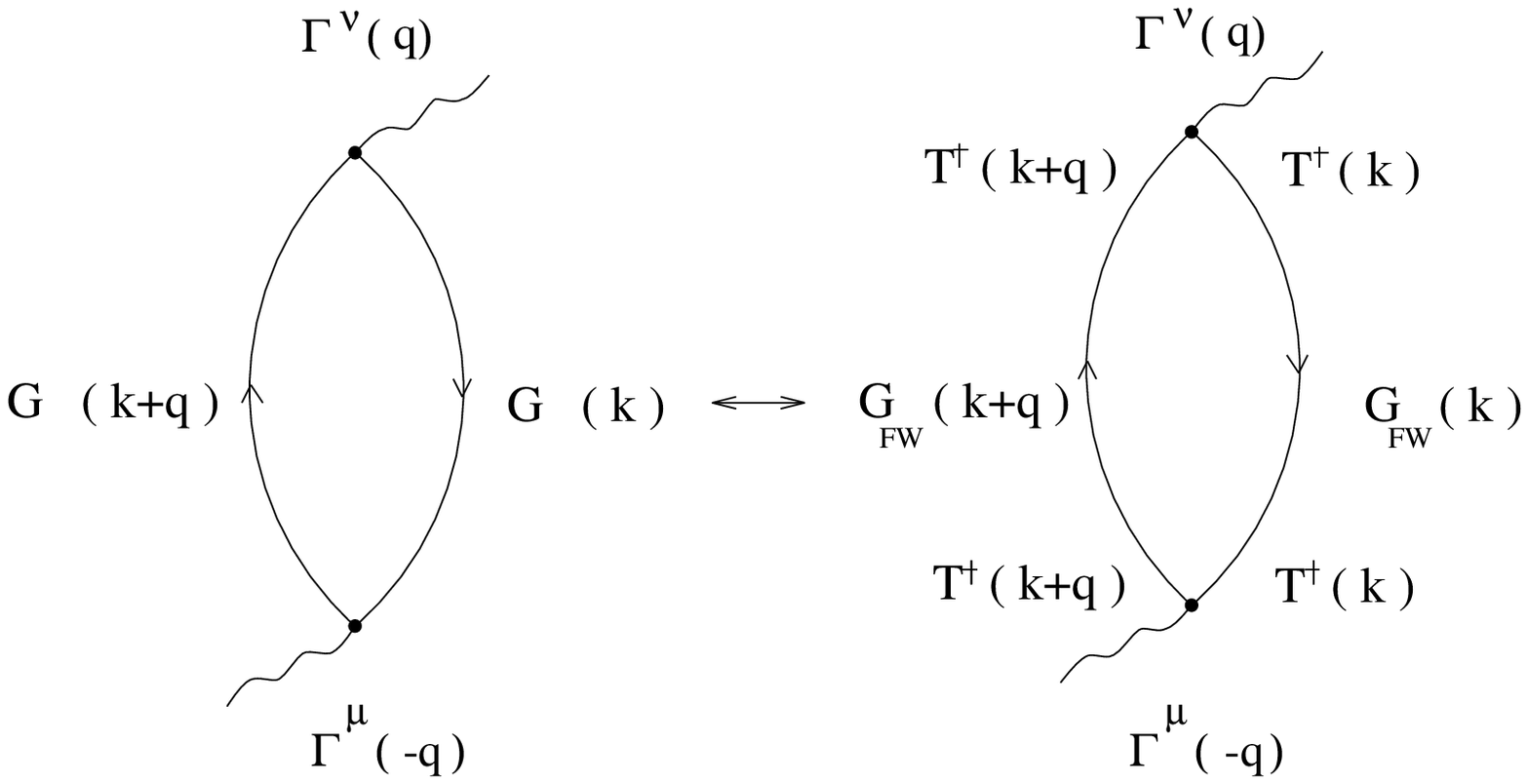}}
\vskip 2mm
\caption{Feynman's rules versus ``Foldy's rules''}
\label{fig:Feyn}
\end{center}
\end{figure}

Fig.~\ref{fig:Feyn} illustrates the passage from the Feynman to the FW rules: 
clearly both the vertices and the Green's functions are changed. The latter are
given by (\ref{eq:fol3}) and the former by
\begin{mathletters}
\label{eq:fol4}
\begin{eqnarray}
\Gamma_{FW}^{\mu} (q) &\equiv&
{\cal T}^\dagger (\bbox{k}+\bbox{q}) \Gamma^{\mu} (q) {\cal T}^\dagger 
(\bbox{k}) \\
\tilde{\Gamma}_{FW}^{\mu} (-q) &\equiv& 
{\cal T}^\dagger (\bbox{k})
\Gamma^{\mu} (-q) {\cal T}^\dagger (\bbox{k}+\bbox{q})  \ \ \ ,
\end{eqnarray}
\end{mathletters}
where $\Gamma^\mu (q) = F_1 (Q^2) \gamma^\mu + i \frac{F_2 (Q^2)}{2 m_N} 
\sigma^{\mu \nu} q^\nu$.

The leading approximation to the space-like charge response amounts to ignoring
all momentum dependence or, equivalently, to allowing for a very large $m_N$
in the vertices
(\ref{eq:fol4}). 
In this scheme one obtains the following FW longitudinal response
function 
\begin{eqnarray}
{\cal R}^{FW}_L (q,\omega) &=& - \frac{V}{\pi} {\rm Im} \Pi_{FW} (q,\omega) = 
\frac{V}{\pi} {\rm Im} \left\{ i \int \frac{d^4k}{(2 \pi)^4} {\rm Tr} \left[
P_+ G_0^{\text{FW}} (k) P_+ G_0^{\text{FW}} (k+q) \right] \right\} \nonumber \\
&=& \frac{3 {\cal N}}{4 m_N \kappa \eta_F^3} \left( \epsilon_F - \Gamma \right) 
\vartheta \left( \epsilon_F - \Gamma \right) \left[ \frac{\kappa^2}{\tau}
\left(1+ \tau +\Delta\right) -\lambda^2 \right]\ \ \ ,
\label{eq:fol5}
\end{eqnarray}
where  the polarization propagator $\Pi_{FW} (q,\omega)$,  calculated in 
terms of the pointlike vertices, has been introduced and where $V$ is the 
(large) volume enclosing the system.

Now, since the only nonzero contributions to $\Pi_{FW}$ come
from ph excitations (space-like region), while those arising from
$p\overline{p}$ excitations are projected out, when integrating 
${\cal R}_{FW} (q,\omega)$ over the energy $\omega$ the closure relation
{\it already applies in the space-like region}, yielding a sum rule which
notably coincides with the NRCSR (\ref{eq:intro1}).
It is of importance to realize that as soon as the zero-momentum
approximation in the vertices is removed, then one
goes beyond the space-like domain and time-like contributions come into play.
Also of relevance is the fact that (\ref{eq:fol5}) cannot be expressed
in terms of the RFG scaling variable (\ref{eq:scal0}).

The complete longitudinal response in the RFG model is expressed
in the FW framework as
\begin{eqnarray}
R_L^{FW} (q,\omega) &=& - \frac{V}{\pi} {\rm Im} \Pi^{00} (q,\omega) 
= V \int_0^{k_F} \frac{d^3 k}{(2 \pi)^3} \delta
(\omega+E_{\bbox{k}}-E_{\bbox{k}+\bbox{q}}) \times \nonumber \\
&\times& \vartheta(|\bbox{k}+\bbox{q}|-k_F) {\rm Tr}\left[
P_+ \Gamma_{FW}^{0} (q) P_+ \tilde{\Gamma}_{FW}^{0} (-q)  \right] \nonumber \\
&=& {\cal R}^{FW}_L (q,\omega) \frac{\frac{\kappa^2}{\tau} \left[ 
G_E^2(\tau) + \Delta W_2(\tau)\right]}{\frac{\kappa^2}{\tau} \left[ 
1+\tau + \Delta \right] - \lambda^2} \ \ \ ,
\label{eq:fol6}
\end{eqnarray}
which, of course, coincides with (\ref{eq:scal1}).
Likewise in the scaling scheme, it is then natural to introduce in the FW 
framework a reduced response according to 
\begin{eqnarray}
r_L^{FW} (\kappa,\lambda) = \frac{R_L^{FW} (\kappa,\lambda)}{H_L^{FW}
 (\kappa,\lambda)} \equiv \frac{1}{{\cal N}} {\cal R}_L^{FW}
(\kappa,\lambda) \ \ \ ,
\label{eq:fol7}
\end{eqnarray}
with the normalizing factor
\begin{eqnarray}
H_L^{FW} (\kappa,\lambda) \equiv {\cal N} 
\frac{\frac{\kappa^2}{\tau} \left[ G_E^2(\tau) + \Delta 
W_2(\tau) \right]}{\frac{\kappa^2}{\tau}  \left[ 1+\tau+\Delta \right] - 
\lambda^2} \ \ \ .
\label{eq:fol8}
\end{eqnarray}
The $r_L^{FW} (\kappa,\lambda)$ defined here then leads automatically to a RCSR
which coincides with the NRCSR. Thus the square roots expressing the energy in
the relativistic propagator (\ref{eq:fol3}) alter the response with respect to
the non-relativistic case, but leave the area under the latter
unchanged.

\subsection{Comparing the two methods} 
\begin{figure}
\begin{center}
\mbox{\epsfig{file=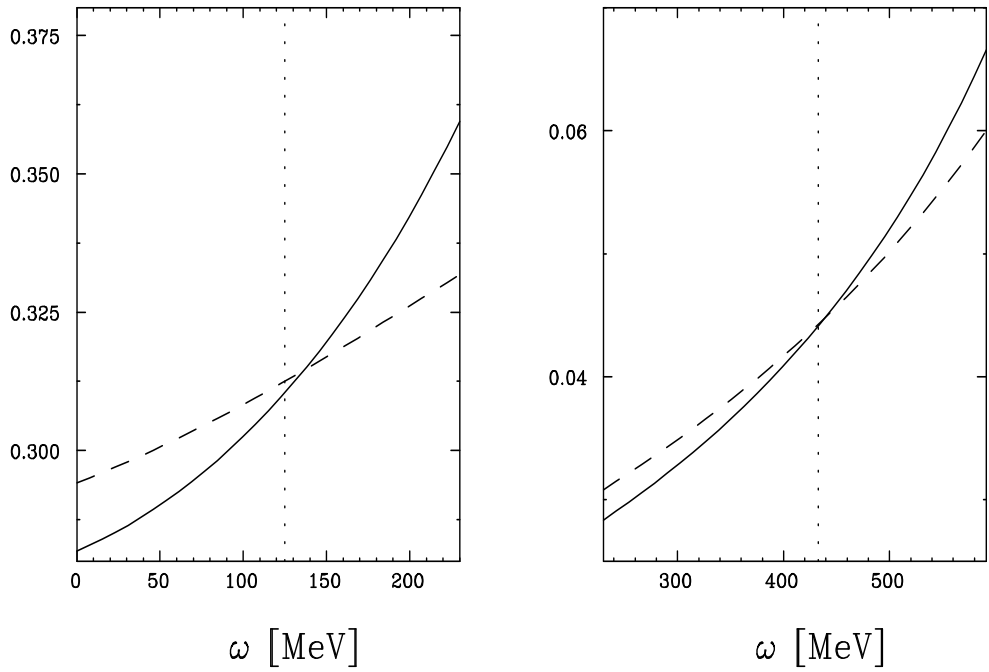,width=14cm,height=12cm}}
\vskip 2mm
\caption{$H_L^{FW}$ (solid) and $H_L^{RFG}$ (dashed) versus $\omega$ over the
response region at $q=500$ MeV/c (panel a) and $q=1$ GeV/c (panel b) for a Fermi
momentum of $k_F=250$ MeV/c. The dotted line shows the position of the
quasielastic peak.}
\label{fig:HFHL}
\end{center}
\end{figure}

\begin{figure}
\begin{center}
\mbox{\epsfig{file=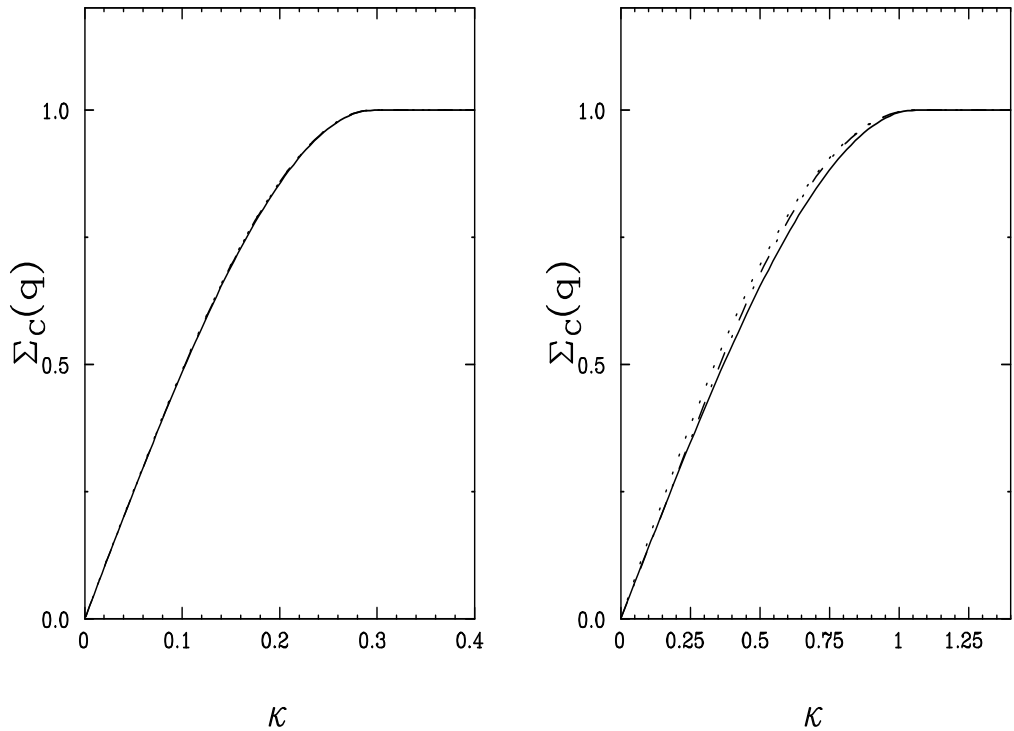,width=14cm,height=12cm}}
\vskip 2mm
\caption{Coulomb sum rule in the scaling framework: exact result numerically
obtained (dotdashed), prediction of formula (\ref{eq:scal3}) (dotted) 
and the NRCSR 
(solid). The three instances are almost indistinguishable at normal nuclear
densities (left panel: $k_F = 250$ MeV/c) whereas at large density (right
panel: $k_F=1000$ MeV/c) relativity appears to
mildly increase the sum rule in the non-Pauli-blocked region.}
\label{fig:scal}
\end{center}
\end{figure}

We now compare $H_L^{FW}(\kappa,\lambda)$ with $H_L^{RFG} (\kappa,\lambda)$ 
in the low-density regime which applies to real nuclei. This is done in 
Fig.~\ref{fig:HFHL}
where the two reducing factors are displayed as a function of $\omega$ for
$q = 500$ MeV/c (panel A) and $q=1$ GeV/c (panel B) respectively. One sees
in the figure that close to the peak the positive $H_L^{RFG}$ and $H_L^{FW}$
essentially coincide, whereas at low (high) frequencies $H_L^{RFG}$ is larger 
(smaller) than $H_L^{FW}$. It is thus clear why both the scaling and FW 
reduced responses
yield the same RCSR in the non-Pauli-blocked regime: indeed they practically
coincide at the peak and their contributions to the sum rule arising from the
edges of the response region add up to the same amount. This compensation, while
almost perfect in the non-Pauli-blocked domain, is not complete in the 
Pauli-blocked one. However the difference is very small, as is apparent in 
fig.~\ref{fig:scal} where the sum rule 
obtained through the scaling approach is plotted at two different Fermi momenta.
The saturation in the non-Pauli-blocked region is evident and, 
moreover, in the Pauli-blocked one, the scaling sum rule is almost
indistinguishable from the 
NRCSR at normal density. Importantly, even at very large densities, the
difference between the two remains quite small.

Let us now compare the reducing factors in leading order of the $\eta_F$
expansion. For this
purpose we recall following \cite{Bar94} that
\begin{eqnarray}
\frac{\partial \psi}{\partial \lambda} &=& \frac{\kappa}{\tau}
\frac{\sqrt{1+\xi_F \psi^2/2}}{\sqrt{2 \xi_F}} \left[ \frac{1+2\lambda+\xi_F
\psi^2}{1+\lambda+\xi_F \psi^2} \right] \\
&=& \frac{\kappa}{\eta_F \tau} \left(\frac{1+2\lambda}{1+\lambda} \right) +
O\left[\eta_F^2\right] \ .
\end{eqnarray}
Therefore one obtains 
\begin{eqnarray}
H_L^{RFG} (\kappa,\lambda)& = &\frac{1+\lambda}{1+2\lambda} {\cal N} G_E^2(\tau)
+{\cal O}[\eta_F^2]\\
& = &\frac{1+\tau}{1+2\tau}\left[1-\frac{1}{1+2\tau}\sqrt{
\frac{\tau}{1+\tau}}\eta_F\psi\right]{\cal N}G_E^2(\tau)+{\cal O}[\eta_F^2]
\end{eqnarray}
and 
\begin{eqnarray}
H_L^{FW} (\kappa,\lambda) & = &\frac{1}{1+\frac{\tau^2}{\kappa^2}}
{\cal N} G_E^2(\tau) +{\cal O}[\eta_F^2]\\
& = &\frac{1+\tau}{1+2\tau}\left[1+\frac{2\tau}{1+2\tau}\sqrt{
\frac{\tau}{1+\tau}}\eta_F\psi\right]{\cal N}G_E^2+(\tau)
{\cal O}[\eta_F^2]\ \ \ .
\end{eqnarray}
to be compared with the De Forest expression \cite{Tdf84}
\begin{equation}
H_L^{\rm DeF}(\kappa,\lambda)=\frac{1+\tau}{1+2\tau}{\cal N}G_E^2(\tau)\ \ \ .
\label{tdfeq}
\end{equation}
One thus sees in the low-density regime  where    $\eta_F\ll 1$
that $H_L^{RFG}$, $H_L^{FW}$ and $H_L^{\rm DeF}$
all coalesce at the peak of the RFG response,
where $\psi=0$, $\lambda=\tau$ and $\kappa^2 = \tau (\tau + 1)$. 
Note also that the differences of ${\cal O}[\eta_F]$ are linear in 
$\psi$ and so tend to cancel when forming integrals over the scaling 
variable with integrands that are symmetric around the quasielastic 
peak.

\section{The time-like region}
As shown by Matsui\cite{Mat83}, the space-like relativistic Coulomb sum 
rule for
pointlike Dirac particles without anomalous magnetic moments goes to $Z/2$ for
large $q$,  showing that for
pointlike nucleons on their mass-shell the particle-antiparticle symmetry is
reflected in the sharing of the Coulomb sum rule at very large momentum
transfers.

Here we consider contributions to the time-like longitudinal response
function arising from the RFG. Thus we write, using the
$p\overline{p}$ polarization propagator,
\begin{eqnarray}
R_L^{tl} (\kappa,\lambda) &\equiv&- \frac{V}{\pi} {\rm Im} \left[
\Pi_{p\overline{p}}^{00} (\kappa,\lambda)|_{k_F=0} - 
\Pi_{p\overline{p}}^{00} (\kappa,\lambda)| \right] = \nonumber \\ 
&=& \frac{3 {\cal N}}{4 m_N \kappa \eta_F^3} 
\left(\tilde{\Gamma}-\zeta_{-}\right) 
\vartheta\left(\tilde{\Gamma}-\zeta_{-}\right) U_L^{tl} (\kappa,\lambda) \ \ \ ,
\label{eq:tl0}
\end{eqnarray}
where, in terms of the time-like electric and magnetic form factors (recall
that the magnitude of the negative $\tau$ always exceeds one in the response
region),
\begin{eqnarray}
&& U_L^{tl} (\kappa,\lambda) \equiv \frac{\kappa^2}{|\tau|} 
\left[G_{E,t}^2(\tau) + W_2(\tau) \Delta_{p\overline{p}} \right] \ \ \ , \\
&& W_2(\tau) = \frac{1}{1+\tau} \left[ G_{E,t}^2 (\tau) + \tau G_{M,t}^2 (\tau) 
\right]  \ \ \ , \\
&& \Delta_{p\overline{p}} \equiv \frac{\tau}{\kappa^2}
\left[ \frac{\left(\zeta_{-}^2+\tilde{\Gamma} \zeta_{-} +\tilde{\Gamma}^2
\right)}{3} - \lambda (\zeta_{-}+\tilde{\Gamma}) + \lambda^2\right] - 
\left(1+\tau\right) \ \ \ , \\ 
&& \zeta_{\pm} \equiv \lambda \pm \kappa \sqrt{1-\frac{1}{|\tau|}} \\ 
\end{eqnarray}
and
\begin{eqnarray}
\tilde{\Gamma} \equiv  min\left[\epsilon_F, \zeta_{+}\right] \ .
\end{eqnarray}
The subtraction in the first line of (\ref{eq:tl0}) also ensures that
the divergences stemming from the Dirac sea are canceled.
As a consequence the integration of $R_L^{tl} (q,\omega)$ over the time-like
region yields the finite contribution\cite{Mat83}
\begin{eqnarray}
\int_{q}^{\infty} \frac{1}{Z} R_L^{tl} (q,\omega) d\omega = - \frac{2}{Z}
\sum_{|\bbox{k}| < k_F} \frac{\left({E_{\bbox{k}+\bbox{q}}-
E_{\bbox{k}}}\right)^2-\bbox{q}^2}{4 E_{\bbox{k}} E_{\bbox{k}+\bbox{q}}}  \ \ \ ,
\end{eqnarray}
which, while vanishing at zero momentum transfer, goes to $1/2$ in the 
large-$q$ limit. 

This result, while of theoretical interest, cannot presently be tested
against experimental data. Indeed, while in the space-like region 
$R_L^{RFG}$ qualitatively, although not quantitatively, accounts for the
experiments, these are lacking in the time-like domain. Furthermore, for
the large energy transfers typical of the time-like sector,
the possibility of exciting the nucleon in the scattering process becomes 
dominant. Finally, even allowing for the possibility of disentangling inelastic
nucleonic processes, our knowledge of the
time-like elastic nucleonic form factors is much poorer than in the 
space-like domain.

\begin{figure}
\begin{center}
\mbox{\epsfig{file=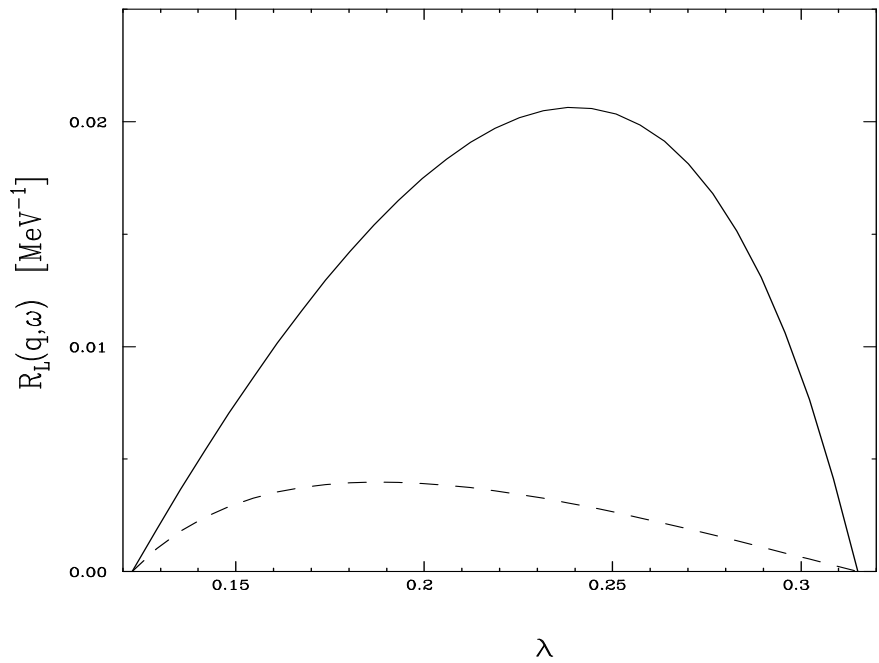,width=12cm,height=10cm}}
\vskip 2mm
\caption{Space-like (solid) and time-like (dashed) longitudinal response for 
Dirac nucleons versus $\lambda$, for $q=1000$ MeV/c, $k_F =250$
MeV/c. The time-like response region has been shifted by -$\epsilon_F$.}
\label{fig:TL}
\end{center}
\end{figure}

Yet, to provide a feeling for 
the time-like physics in the simple point-nucleon
RFG model, in Fig.~\ref{fig:TL} the space-like and time-like longitudinal 
response functions are plotted as functions of $\lambda$
for $q = 1000$ MeV/c and $k_F = 250$ MeV/c. 
Note that the boundaries of the time-like response are
the same as the space-like ones but for a shift of $\epsilon_F$ (corresponding
in $\omega$ essentially to $2 m_N$).
Accordingly the time-like response $R_L^{tl}$ is shown in the figure for  a
range displaced by the Fermi energy to make the comparison with $R_L$ easier.
Note that the off-set of the maximum of  $R_L^{tl}$
with respect to that of $R_L$ is no longer given by $\lambda_0$ and that 
in general the two responses  have different
shapes and norms.

\begin{figure}
\begin{center}
\mbox{\epsfig{file=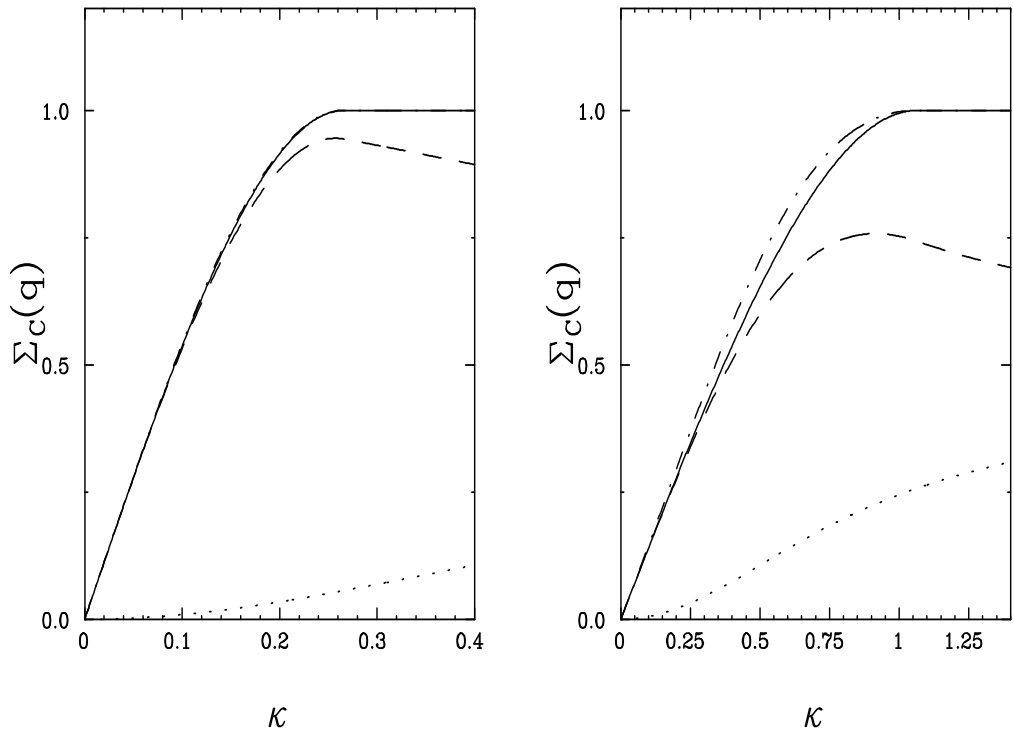,width=14cm,height=12cm}}
\vskip 2mm
\caption{Coulomb sum rule for pointlike nucleons without anomalous magnetic
moments: NRCSR (solid), $ph$ excitation (dashed), $p\overline{p}$ 
excitations (dotted) and $ph+p\overline{p}$ (dotdashed).
Left panel: $k_F = 250$ MeV/c, right panel: $k_F = 1000$ MeV/c.
Observe that only at large $k_F$ is
the difference between the NRCSR and the total
relativistic sum rule perceptible. Here relativity, as in the scaling
framework, appears to increase the sum rule somewhat.
Observe also the growth of the $p\overline{p}$ contribution with $k_F$.}
\label{fig:Coul}
\end{center}
\end{figure}

Likewise for purpose of illustration, we display in Fig.~\ref{fig:Coul} the 
Coulomb sum rule for the RFG of Dirac nucleons, 
including the contribution stemming from $p\overline{p}$ excitations. 
The sum rule, defined through a frequency integral spanning both the space-like
and the time-like regions and suitably renormalized, reads\cite{Mat83}
\begin{eqnarray}
\Sigma_C (q) = 1 - \frac{3}{4 \pi k_F^3} \int d^3k
\frac{(E_{\bbox{k}+\bbox{q}}+E_{\bbox{k}})^2-\bbox{q}^2}{4
E_{\bbox{k}} E_{\bbox{k}+\bbox{q}}} \vartheta\left(k_F-k\right)
\vartheta\left(k_F - |\bbox{k}+\bbox{q}|\right) \ \ \ .
\label{eq:tl1}
\end{eqnarray} 
One sees from the figure that $\Sigma_C$ saturates when the Pauli correlations
are no longer operative (i.e. for $\kappa>0.27$ at $k_F=250$ MeV/c and for
$\kappa > 1.06$ at $k_F=1000$ MeV/c) and that, in the Pauli-blocked region,
the difference between the
relativistic and the non-relativistic Coulomb sum rule is barely perceptible
except for very large densities, since (\ref{eq:tl1}) correctly reduces to 
the NRCSR in the limit $k_F < m_N$. Moreover the $p\overline{p}$ 
contribution starts to be substantially felt only at large $k_F$.

\section{Beyond the step-function momentum distribution}
With the idea of exploring the impact of nucleon-nucleon correlations
on the Coulomb sum rule let us insert into the RFG model 
momentum distributions that are more general than the step-function. 
For this purpose,
following \cite{Don89} and exploiting (\ref{eq:simmetria}), we first 
observe that $R_L$
can be expressed as follows\footnote{In this
expression the energy of the struck nucleon is simply assumed to be the RFG
energy $E_{\bbox{p}}$. Thus off-shell effects in the e.m. 
vertices, which are not easy to predict since we lack 
a fundamental theory, have been ignored.}
\begin{eqnarray}
R_L^{RFG} (\kappa,\lambda) &=& \frac{1}{4 \pi m_N} \left\{ \int d^3\eta n(\eta) 
\frac{X_L^{RFG}(\eta,\kappa,\lambda)}{X_L^{FW}(\eta,\kappa,\lambda)}
\delta\left(2 \lambda + \sqrt{1+\eta^2}-\sqrt{1+(\bbox{\eta}+2\bbox{\kappa})
^2}\right) - \right. \nonumber \\
&-& \left. \int d^3\eta n(\eta) 
\frac{X_L^{RFG}(\eta,\kappa,-\lambda)}{X_L^{FW}(\eta,\kappa,-\lambda)}
\delta\left(-2\lambda + \sqrt{1+\eta^2}-\sqrt{1+(\bbox{\eta}+2 \bbox{\kappa})^2}
\right)\right\} \\
&=& \frac{1}{2 m_N} \left\{ \int_{y_{-}}^{\infty} d\eta
n(\eta) X_L^{RFG}
(\eta,\kappa,\lambda)- \int_{y_{+}}^{\infty} d\eta n(\eta) 
X_L^{RFG}(\eta,\kappa,-\lambda)
\right\} \ \ \ ,
\label{eq:mom2}
\end{eqnarray} 
where 
\begin{eqnarray}
y_{\pm} \equiv | \kappa \pm \lambda \sqrt{1+1/\tau} | \ \ \ , \nonumber  
\end{eqnarray} 
\begin{eqnarray}
X_L^{RFG}(\eta,\kappa,\lambda) &\equiv& \frac{\eta}{2 \kappa \epsilon} \left\{
\left(\epsilon+\lambda\right)^2 \frac{1}{1+\tau}\left[G_E^2(\tau)+\tau 
G_M^2(\tau)\right] - \kappa^2 G_M^2(\tau) \right\} = \nonumber \\
&=& \frac{\eta}{2 \kappa \epsilon} \frac{\kappa^2}{\tau} \left(G_E^2(\tau)
+ W_2(\tau) \eta_{\perp}^2(\lambda,\kappa) \right) \ \ \ ,  \\
X_L^{FW}(\eta,\kappa,\lambda) &=& \frac{\eta}{2 \kappa} (\epsilon + 2 \lambda) 
\label{eq:mom3}
\end{eqnarray} 
and the momentum distribution $n (\eta)$ is normalized according to 
\begin{eqnarray}
\int_0^{\infty} \eta^2 n(\eta) d\eta = {\cal N} \ \ \  
\end{eqnarray} 
separately for protons and neutrons.
In the same limit as in (\ref{eq:fol5}) $X_L \rightarrow
X_L^{FW}$, which explains the notation of the LHS of (\ref{eq:mom3}). 
The response one obtains is then just a generalization of the FW one, namely
\begin{eqnarray}
{\cal R}_{FW} (\kappa,\lambda) = 
\frac{1}{2 m_N {\cal N}} \left\{ \int_{y_{-}}^{\infty} d\eta n(\eta) 
X_L^{FW}(\eta,\kappa,\lambda) - \int_{y_{+}}^{\infty} d\eta n(\eta) 
X_L^{FW}(\eta,\kappa,-\lambda) \right\} \ \ \ ,
\label{eq:mom4}
\end{eqnarray} 
which, setting $n (\eta) = \frac{3 {\cal N}}{\eta_F^3}
\theta(\eta_F-\eta)$, yields (\ref{eq:fol5}). 
 
For the above response the easily obtained Coulomb sum rule (the
tilde is to remind us that a generic momentum distribution is employed) reads
\begin{eqnarray}
\tilde{\Sigma}_C &\equiv& \int_0^{q} {\cal R}_{FW} (q,\omega) d\omega = 
\frac{1}{4 \pi {\cal N}} \int d^3\eta
n(\eta) \left[ \vartheta\left(\cos\theta +\frac{\kappa}{\eta}\right) - 
\vartheta\left(-\cos\theta -\frac{\kappa}{\eta}\right) \right] = \nonumber \\
&=& 1 + \frac{1}{\cal N} \int_{\kappa}^{\infty} d\eta \eta^2 n(\eta)
\left(\frac{\kappa}{\eta}-1\right)\ \ \ ,
\label{eq:mom5}
\end{eqnarray} 
which is bound to saturate in the large momentum limit but for pathological
momentum distributions. By expanding (\ref{eq:mom5})  around $\kappa = 0$ 
one obtains 
\begin{eqnarray}
\tilde{\Sigma}_C \approx \frac{\kappa}{\cal N} \int_{0}^{\infty} d\eta
\eta n(\eta) - \frac{n(0)}{6 {\cal N}} \kappa^3 + O(\kappa^4) \ \ \ ,
\label{eq:mom6}
\end{eqnarray} 
which has the behaviour typical of a sum rule of an infinite system.

From (\ref{eq:mom6}) the  NRCSR for the FG is recovered by
employing a step-function momentum distribution. On the other hand for a
general $n(\eta)$, one can still have 
the term linear in $\kappa$ in (\ref{eq:mom6}) identical to the one appearing 
in the NRCSR for the FG by setting
\begin{eqnarray}
\eta_F = \frac{3}{2} \frac{\cal N}{\int_0^{\infty} d\eta \eta n(\eta)} \ \ \ , 
\label{eq:mom7}
\end{eqnarray} 
which can be exploited for relating $\eta_F$ to a
finite nucleus momentum distribution.
In this connection we recall that in ref.\cite{Cen96} two procedures have been
suggested to determine the Fermi momentum in a way that brings the RFG as 
close as possible to a real nucleus, one related to the nuclear momentum distribution
and the other
to the half width of the longitudinal response. These yield for $^{16}$O
the values $k_F=1.039$ fm$^{-1}$ and $k_F=1.22$ fm$^{-1}$ respectively.
Here (\ref{eq:mom7}) yields  $k_F=1.1$ fm$^{-1}$ for $^{16}$O,
which lies in between the above quoted values.

To define a reducing factor in the presence of a
generic momentum distribution it helps to
notice that $\Delta$, the quadratic transverse momentum distribution, for any
$n(\eta)$ would read 
\begin{eqnarray}
\tilde{\Delta} = \frac{\int d^3\eta n(\eta) 
\frac{1}{\epsilon (\epsilon+2\lambda)} \eta_{\perp}^2(\kappa,\lambda)}{\int 
d^3\eta n(\eta) \frac{1}{\epsilon (\epsilon+2\lambda)}} = 
\frac{\int_{y_{-}}^{\infty} d\eta \frac{\eta}{\epsilon} 
\eta_{\perp}^2(\kappa,\lambda) n(\eta)}{\int_{y_{-}}^{\infty} d\eta 
\frac{\eta}{\epsilon} n(\eta)} \ \ \ ,
\label{eq:mom8}
\end{eqnarray} 
and would reduce to (\ref{eq:scal2}) for a step-function momentum 
distribution.
Accordingly, in a FW-inspired framework, one might introduce
\begin{eqnarray}
\tilde{r}_L^{FW} 
(\kappa,\lambda)\equiv\frac{R_L(\kappa,\lambda)}{\tilde{H}_L^{FW}
(\kappa,\lambda)} \ \ \ ,
\end{eqnarray} 
with
\begin{eqnarray}
{\tilde{H}_L^{FW}(\kappa,\lambda)} = {\cal N} \frac{\frac{\kappa^2}{\tau} 
\left[ G_E^2(\tau) + \tilde{\Delta} W_2(\tau) \right]}
{\frac{\kappa^2}{\tau}  \left[ 1+\tau+\tilde{\Delta} \right] - 
\lambda^2 } \ \ \ 
\label{eq:mom9}
\end{eqnarray} 
and likewise, in the scaling scheme, one would use $\tilde{H}_{L}^{RFG}(\kappa,
\lambda)$ as given by (\ref{eq:cenni}), but with $\tilde{\Delta}$ replacing
$\Delta$. 

To get a feeling for the impact of using a generalized
momentum distribution on the Coulomb sum
rule we take as an example the simple harmonic oscillator
description of $^{16}$O. The associated momentum distribution, to be {\it ad
hoc} inserted in the RFG model, reads
\begin{eqnarray}
n(\eta)=\frac{8}{\sqrt{\pi}\eta_0^3}\left(1+2(\eta/\eta_0)^2\right)
e^{-(\eta/\eta_0)^2} \ \ \ 
\label{eq:mom10}
\end{eqnarray} 
with $\eta_0 = \sqrt{\omega_0/m_N}$.

\begin{figure}
\begin{center}
\mbox{\epsfig{file=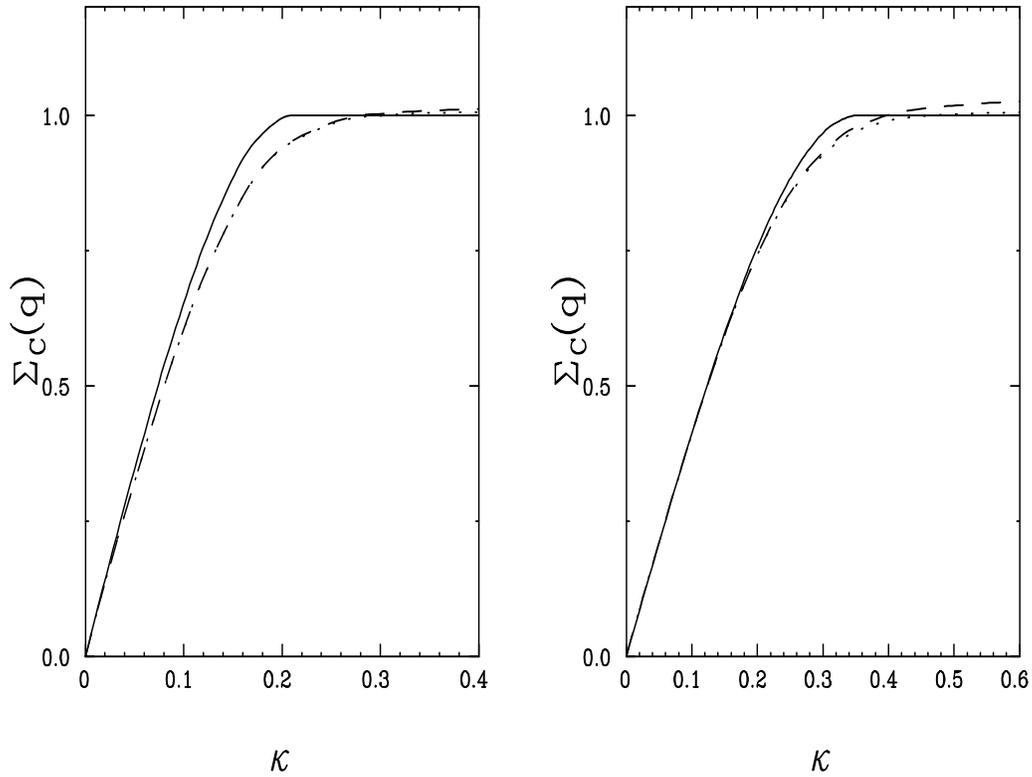,width=14cm,height=12cm}}
\vskip 2mm
\caption{The Coulomb sum rule for the RFG as given by the Foldy
procedure with  a shell model momentum distribution. 
In both panels we display the NRCSR for a Fermi momentum given by 
$\eta_F=\protect\sqrt{\pi} \eta_0$ (solid line)
and the sum rule corresponding to the 
momentum distribution (\protect\ref{eq:mom10}) with $\eta_0=0.13$
(left panel) and $\eta_0=0.2$ (right panel). The dashed and dotted lines refer
to the sum rules one obtains with the longitudinal 
response reduced by $H_L^{FW}$ and
by $\tilde{H}_L^{FW}$ respectively. It is evident that for large $\eta_0$ the
reducing factor $\tilde{H}_L^{FW}$ is necessary 
to obtain the correct saturation.}
\label{fig:sum}
\end{center}
\end{figure}

\begin{figure}
\begin{center}
\mbox{\epsfig{file=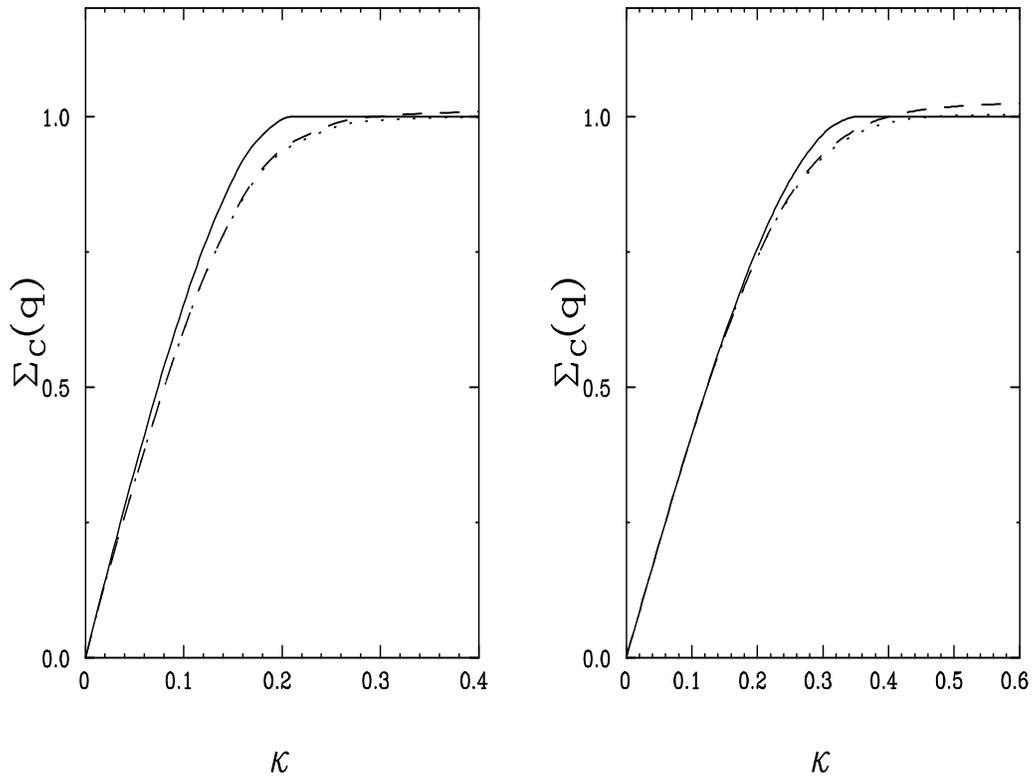,width=14cm,height=12cm}}
\vskip 2mm
\caption{The Coulomb sum rule for the RFG as obtained in the 
scaling framework with a shell model momentum distribution. 
As in the previous figure the solid line is the NRCSR for 
a Fermi momentum given by $\eta_F=\protect\sqrt{\pi} \eta_0$ and the sum rule 
for the momentum distribution (\protect\ref{eq:mom10}) with $\eta_0=0.13$ 
(left panel) and $\eta_0=0.2$ (right panel) are displayed as well.
The dashed and dotted lines refer to the sum rules for the longitudinal 
response reduced by $H_F^{FW}$ and by $\tilde{H}_F{FW}$ respectively. 
It is again apparent the necessity of introducing $\tilde{H}_F{FW}$ when 
$\eta_0$ is large.} 
\label{fig:sum1}
\end{center}
\end{figure}

In Fig.~\ref{fig:sum} and Fig.~\ref{fig:sum1} we display a few versions of
$\tilde\Sigma_C(q)$
for the momentum distribution (\ref{eq:mom10}) as obtained in the FW and
scaling scheme respectively. In the left panels of the 
figures we have set $\eta_0 = 0.13$, which follows from the formula 
$\hbar \omega_0 =41/A^{1/3}$ MeV for the harmonic oscillator
frequency. In the right panels the larger value $\eta_0=0.2$ has 
instead been chosen. It is noteworthy that, in order to achieve
saturation
for large values of $\eta_0$, namely for extended momentum
distributions, the reduction factors should be
calculated utilizing (\ref{eq:mom8}) rather than  (\ref{eq:scal2}).

In conclusion we provide the analytic expression for the expansion of 
(\ref{eq:mom6}) in the case of the momentum distribution (\ref{eq:mom10}). It
reads
\begin{eqnarray}
\tilde{\Sigma}_C &\approx& \frac{3 \kappa}{2 \sqrt{\pi} \eta_0} -
\frac{\kappa^3}{6 \sqrt{\pi} \eta_0^3} + O(\kappa^4) \ 
\end{eqnarray} 
and therefore (\ref{eq:mom7}) in this case simply becomes
\begin{eqnarray}
\eta_F = \sqrt{\pi} \eta_0 \ \ \ .
\end{eqnarray} 

\section{Energy-weighted sum rules}
In this section we consider the energy-weighted (EWSR) and the 
squared-energy-weighted sum rules. For this purpose we again exploit 
(\ref{eq:simmetria}) 
which allows us to express the longitudinal Pauli-blocked response function in
terms of the non-Pauli-blocked one. Accordingly the EWSR may be cast in the 
following form
\begin{eqnarray}
\Sigma_{E} &=& \int_0^{\infty} d\omega \omega r_L (q,\omega) =
\int_0^{\infty} d\omega \omega \left[ r_L^{nb} (q,\omega) - r_L^{nb} (q,-\omega)
\right] = \nonumber \\
&=& \int_{-\infty}^{\infty} d\omega \omega r_L^{nb} (q,\omega) \ \ \ .
\label{eq:ew1}
\end{eqnarray} 
From the above expression it follows that the passage from
the Pauli-blocked regime to the non-Pauli-blocked one {\it leaves 
unaffected the functional form of $\Sigma_{E}$}, unlike in the case of
$\Sigma_C$. More generally, the above result
holds for all the sum rules with odd powers of $\omega$.

In the non-relativistic case, where the response function is symmetric with 
respect to its peak, (\ref{eq:ew1}) just selects  the peak itself and
accordingly one has
\begin{eqnarray}
\Sigma_{E} = \frac{q^2}{2 m_N} \ \ \ .
\label{eq:ew1bis}
\end{eqnarray} 
In the relativistic case, where this symmetry is lost and thus the peak of the
response does not occur in correspondence with the midpoint of the response 
region, the EWSR is given by the relativistic extension of (\ref{eq:ew1bis}), 
namely $|Q^2|/2 m_N$, plus medium dependent corrections. 

To ascertain whether the latter are larger in the FW or in the scaling 
approach we quote the expressions for the the energy-weighted and the 
squared-energy-weighted sum rules
(the latter only valid for $\kappa > \eta_F$) up to 
second order in the expansion in $\eta_F$ as obtained in the scaling approach,
namely\cite{Cen96}
\begin{mathletters}
\begin{eqnarray}
\lefteqn{\Xi^{(1)} = \int_0^{\infty} d\omega \lambda r_L (\psi,\omega) 
}
\label{5.3a}\\
&&=\lambda_{0} + \frac{\eta_F^2}{20 (1+4\kappa^2)^{3/2}} \left[ 1 -
(1+4\kappa^2)^{3/2} \right]  + O\left[\eta_F^3\right]
\nonumber
\end{eqnarray}
\begin{eqnarray}
\lefteqn{\Xi^{(2)} = \int_0^{\infty} d\omega \lambda^2 r_L 
(\psi,\omega)}
\label{5.3b}\\
&&=\lambda_{0}^2 + \frac{\eta_F^2}{10 (1+4\kappa^2)^{3/2}} \left[
(1+4\kappa^2)^{3/2} - 1 - 4 \kappa^2 - 8 \kappa^4 \right] +
O\left[\eta_F^3\right]  \ \ \ ,
\nonumber
\end{eqnarray}
\end{mathletters}
and in the FW approximation, namely
\begin{mathletters}
\begin{eqnarray}
\lefteqn{\Xi^{(1)} = \int_0^{\infty} d\omega \lambda r_L (q,\omega) }
\label{5.4a}\\
&&= \lambda_{0} + \frac{3 \eta_F^2}{20 (1+4\kappa^2)^{3/2}} \left[ 1 
 - (1+4\kappa^2)^{3/2} + \frac{8}{3}
\kappa^2 \right]  + O\left[\eta_F^3\right] 
\nonumber
\end{eqnarray}
\begin{eqnarray}
\lefteqn{\Xi^{(2)} = \int_0^{\infty} d\omega \lambda^2 r_L (q,\omega)}
\label{5.4b}\\
&& =\lambda_{0}^2 + \frac{3 \eta_F^2}{10 (1+4\kappa^2)^{3/2}} \left[
(1+4\kappa^2)^{3/2} - 1 - \frac{16}{3} \kappa^2 - 8 \kappa^4 \right] +
O\left[\eta_F^3\right]  \ \ \ .
\nonumber
\end{eqnarray}
\end{mathletters}
Interestingly in both cases the variance turns out to be
\begin{eqnarray}
\sigma = \sqrt{\Xi^{(2)} - (\Xi^{(1)})^2} = \frac{\kappa \eta_F}
{\sqrt{5 (1+4\kappa^2)}} + O\left[ \eta_F^3 \right] \ \ \ .
\end{eqnarray} 

\begin{figure}
\begin{center}
\mbox{\epsfig{file=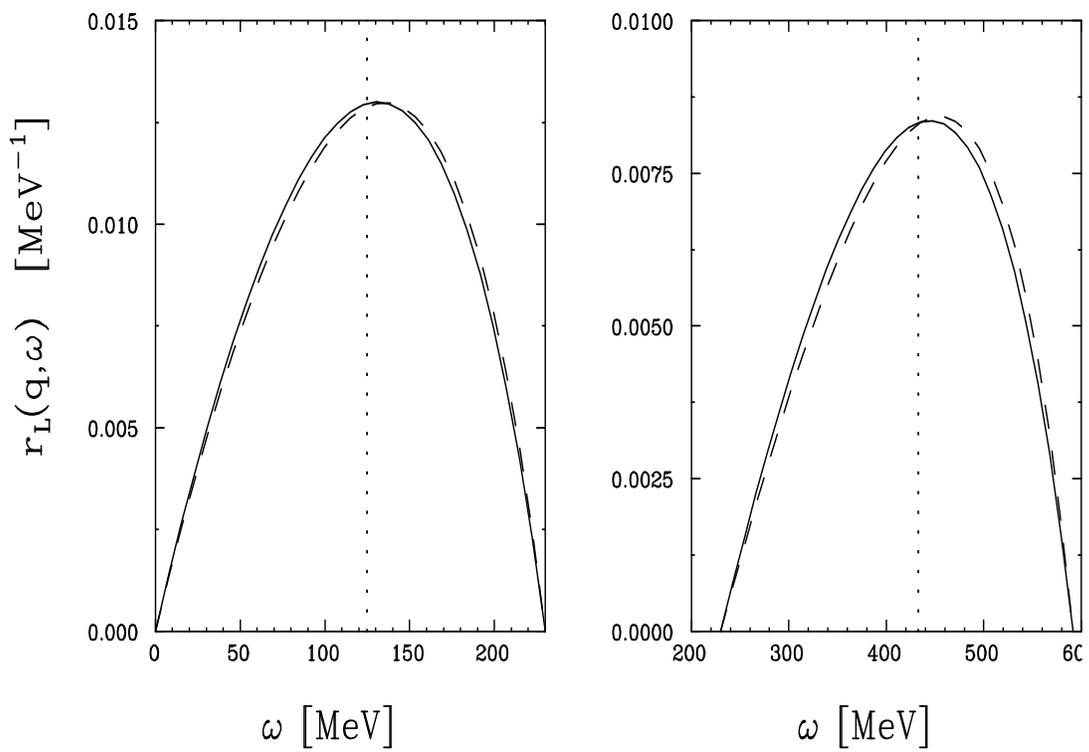,width=14cm,height=12cm}}
\vskip 2mm
\caption{The reduced responses for the FW method (solid) and for the scaling
one (dashed) for $k_F = 250$ MeV/c. Left panel: $q = 500$ MeV/c, right panel:
$q = 1000$ MeV/c.}
\label{fig:figure8}
\end{center}
\end{figure}

We thus reach the conclusion that the FW reduced response is somewhat softened
with respect to the scaling one and that both procedures lead 
to the same variance (or width of the reduced response at half-height) to 
$O\left[ \eta_F^3 \right]$.
The softening is, however, quite modest and can only be appreciated at large
momentum transfers for normal densities (see Fig.~\ref{fig:figure8}). It becomes
pronounced at very large $k_F$, which of course has no physical significance,
where also the widths become much different. 
\newpage

\section{Conclusions}
In order to fulfill a relativistic Coulomb sum rule the charge response 
should be normalized through a factor devised to divide out the physics of the 
nucleon to the extent that is possible. The latter, however, is 
presently not calculable from 
a fundamental theory and is accordingly expressed through phenomenological
form factors parametrized in the space-like region. Moreover the time-like
physics is inaccessible with presently available experimental
facilities, and hence leads us to
the necessity of constructing space-like relativistic sum rules.  

In this paper we have compared, in the framework of the RFG model, two different
methods for setting up the normalizing factor that yields a space-like 
relativistic Coulomb sum rule with the same basic feature provided by
the NRCSR, namely 
the saturation in the non-Pauli-blocked regime.
The first method, established a few years ago by Alberico et al.\cite{Alb88}, 
exploits the scaling behavior of the reduced longitudinal
response function at large momentum transfers, whereas the second makes use
of the FW transformation, which allows one to
exhaust the completeness of the states of the RFG hamiltonian 
in the space-like domain alone.

The two methods provide reduced responses that are
modestly different only at large
momentum transfers (where the FW response turns out to be slightly
softened with respect to the scaling one), have the same variance (up to terms
proportional to $\eta_F^2$) and fulfill the Coulomb sum
rule to any order in $\eta_F$ in the Pauli-unblocked region.
Also in the presence of Pauli blocking, the two sum rules are
very close to each other in the range of $k_F$
appropriate for nuclei, the RCSR
of the FW method coinciding with the well-known non-relativistic sum 
rule.

We thus conclude that it is indeed possible, in the framework of the
RFG,  to obtain a space-like RCSR with the correct 
saturating behaviour for {\it any}
$k_F$. Actually, as we have seen, the procedure for achieving this is not 
unique; but it is gratifying that the two methods we have explored yield reduced
responses that are close to each other over a large span
of momentum transfers at normal nuclear densities. It might be worth
observing, however, that as the density and
hence the magnetic contribution to the charge response (proportional to the 
transverse nucleon's momentum) grow, then the two reduced responses 
referred to above start to
differ more and more. And yet the one obtained for example 
in the scaling approach
still fulfills the sum rule at very large $k_F$ where it becomes
extremely distorted. 

The question may then be asked whether a Coulomb sum rule can still be defined
in the presence of strong correlations among nucleons yielding distributions
extending to momenta larger than those allowed by a step-function. 
By inserting a simple model for such distributions in the RFG framework
we have shown that the Coulomb sum rule still exists, providing
the extended momentum distribution is inserted as well into the normalizing
factors $H_L$, originally defined in terms of the pure
$\theta$-function.
Whether or not our finding, obtained  for a $k_F$ corresponding to 
normal nuclear
density, stays valid at larger $k_F$ remains to be explored. 

Furthermore, while the nucleons in the RFG are on their mass shell, they move 
off it when correlations come into
play. We have not accounted for this physics in the present work. However we 
feel supported by what we view as an important finding from ref.\cite{Cen96}
where in the framework of a hybrid model it has been shown that the
structure of the normalizing factor for the RFG as provided by the scaling
approach is not altered by the off-shellness of the nucleons if an appropriate
shift of the energy transfer $\omega$ is performed.

In this respect the modification of the normalizing factor introduced here 
to reflect the effect of an extended momentum distribution 
complements the one of ref.
\cite{Cen96}, where it reflects the nucleon's off-shellness. 
Still to be performed is a combination of the two approaches in a scheme
encompassing both the nucleon's confinement 
(hybrid model) and realistic nucleon-nucleon correlations 
(extended momentum distribution).

\newpage

\newpage

\newpage

\newpage

\newpage

\newpage

\newpage

\newpage

\end{document}